# THERMODYNAMIC SCALING OF THE VISCOSITY OF VAN DER WAALS, H-BONDED, AND IONIC LIQUIDS

C.M. Roland
*Naval Research Laboratory, Chemistry Division, Code 6120, Washington DC 20375-5342*

S. Bair
*Georgia Institute of Technology, Center for High Pressure Rheology, Atlanta, GA 30332-0405*

R. Casalini
*George Mason University, Chemistry Department, Fairfax VA 22030 and
Naval Research Laboratory, Chemistry Division, Code 6120, Washington DC 20375-5342*

(June 31, 2006)

Viscosities, $\eta$, and their temperature, $T$, and volume, $V$, dependences are reported for 7 molecular liquids and polymers. In combination with literature viscosity data for 5 other liquids, we show that the superpositioning of relaxation times for various glass-forming materials when expressed as a function of $TV^\gamma$, where the exponent $\gamma$ is a material constant, can be extended to the viscosity. The latter is usually measured to higher temperatures than the corresponding relaxation times, demonstrating the validity of the thermodynamic scaling throughout the supercooled and higher $T$ regimes. The value of $\gamma$ for a given liquid principally reflects the magnitude of the intermolecular forces (e.g., steepness of the repulsive potential); thus, we find decreasing $\gamma$ in going from van der Waals fluids to ionic liquids. For strongly H-bonded materials, such as low molecular weight polypropylene glycol and water, the superpositioning fails, due to the non-trivial change of chemical structure (degree of H-bonding) with thermodynamic conditions.

**INTRODUCTION**

Identifying the intermolecular potential governing the interactions in liquids is central to understanding their thermodynamic and other properties. This statement is especially pertinent for the supercooled regime, in which the mobility of viscous liquids is governed by (infrequent) jumps over barriers large in comparison to the available thermal energy. However, due to the cooperativity inherent to vitrifying liquids, each molecule is involved with competing interactions with many neighbors, and an interpretation requires consideration of the complex topography of the multi-dimensional potential energy landscape [1,2,3,4,5,6]. One simplification is to approximate the interactions with a spherically symmetric, pairwise additive, repulsive potential, [7,8,9]

$$U(r) \sim \left(\frac{\sigma}{r}\right)^{3\gamma} \quad (1)$$

where $r$ is the intermolecular separation, and $\sigma$ and $\gamma$ are material-specific constants related respectively to the molecular size and the steepness of the potential. There is no explicit consideration of attractive forces, since solidification is primarily driven by the repulsive forces [10,11,12].

Eq.(1) implies that the product of the glass transition temperature, $T_g$, times the specific volume, $V$, raised to the power of $\gamma$, will be a constant. This was verified in molecular dynamic simulations (mds) of a Lennard-Jones (L-J) fluid ($3\gamma = 12$), for which the quantity $T_g V_g^4$ ($V_g$ is specific volume at the glass transition) was found to be independent of pressure [13,14]. A similar result was obtained from quasielastic neutron scattering on OTP by Tolle et al. [15] for the dynamic crossover temperature of mode coupling theory. Subsequently, Hollander and Prins [16] determined by NMR measurements at high pressures that $T_g V_g^2$ was constant for atactic polypropylene, the smaller value of $\gamma$ reflecting the softer potential for the polymer (as discussed below).

The constancy of the product $T_g V_g^\gamma$ follows from the fact that the relaxation time at $T_g$ is constant [17]. Thus, the invariance of $T_g V_g^\gamma$ can be generalized to all relaxation times by superimposing $\tau(T,P)$ vs. $TV^\gamma$.; that is, $\tau$ is a unique function of $TV^\gamma$

$$\tau(T,V) = \Im(TV^\gamma) \quad (2)$$

as first shown for OTP with $\gamma =4$ [18,19,20], and subsequently for a wide range of glass-forming liquids and polymers ($0.13 \leq \gamma \leq 8.5$) by dielectric spectroscopy [21,22,23,24,25,26,27], light scattering [28], and mds [29,30]. Eq.(2) also applies to polymer blends, with each component exhibiting a unique value of $\gamma$ [31,32]. Furthermore, since the scaling exponent can be deduced from PVT measurements [33] or identified with the Grüneisen parameter [34,35], the scaling relation (eq.(2)) enables the volume and pressure dependences of $\tau$ to be determined from relaxation measurements made only at atmospheric pressure. The parameter $\gamma$ has been found to correlate with the steepness index (fragility [36]) at atmospheric pressure [37] and ultimately with the general T-dependence [38].

In addition to providing a means to characterize relaxation times of glass-formers, the appeal of the scaling is its putative connection to the intermolecular potential. The form of eq.(1) is expected to apply to van der Waals or other liquids lacking strong specific interactions; nevertheless hydrogen-bonded liquids such as sorbitol [21] and glycerol [26,27] conform to eq.(2). It is surprising that the relaxation times for a liquid whose chemical structure is sensitive to temperature (and perhaps pressure [39,40,41,42]) would superpose for a constant value of $\gamma$. We examine this further herein, considering data for polypropylene glycol and water, nonspherical molecules with extensive, directional hydrogen bonding.

Along these same lines, it is of interest to determine if the relaxation properties of ionic liquids can be described using eq.(2). Ionic liquids are molten organic salts, usually with low (< 100°C) melting points. They find increasing applications, for example as substitutes for traditional solvents, due to their good dissolution properties, high thermal stability, and negligible vapor pressure [43]. The interactions in these polar, non-coordinating liquids are complex, entailing electrostatic as well as van der Waals bonding. Viscosity, $\eta$, and PVT measurements have been carried out at elevated pressures for three ionic liquids [44,45], and these data allow assessment of the applicability of the thermodynamic scaling. The viscosity is an especially important property of ionic liquids since it governs the diffusivity and thus rate of chemical reactions carried out in an ionic liquid medium. To date, viscosity scaling has only been applied to results for OTP [24] and glycerol [28], with the obtained exponents close to the $\gamma$ from relaxation measurements [18,26,27]. Approaching the glass transition, $\eta$ of a liquid increases by many orders of magnitude over a small temperature range, mirroring the behavior of the structural relaxation time. According to the Einstein-Debye relation [46,47],

$$\eta = \left(\frac{kT}{v_m}\right)\tau \qquad (3)$$

where $v_m$ is the molecular volume. This equation was derived originally for Brownian particles, which are larger than molecules. An analogous relation is the Maxwell equation [48]

$$\eta = G_\infty \tau \qquad (4)$$

where $G_\infty$ is the infinite-frequency (glassy) shear modulus of the liquid. These equations are commonly used both for liquids and for probe molecules dispersed in a fluid. As the glass transition temperature is approached from above, eq.(4) is found to underestimate the viscosity [49,50,51], a phenomenon referred to as decoupling. Herein we analyze results, primarily viscosity data, for various liquids. These range from simple organic molecules and higher molecular weight compounds (including for example a perfluorinated polyalkylether used in applications such as computer hard drives, uranium enrichment centrifuges, and jet engines), to associated liquids and molten salts. From these data, we determine the general applicability of eq.(2).

**EXPERIMENTAL**

Viscosities were measured using three falling-ball viscometers [52,53]. PVT data were obtained with either a metal bellows piezometer [54] or a commercial Gnomix instrument which uses a similar bellows with the sample confined in mercury [55]. The squalane (hexamethyltetracosane) was obtained Sigma-Aldrich and had a molecular weight $M_w$ = 422.8 g/mol (7 repeat units). The mineral oil was LVI 260 from Shell Lubricants, with $M_w$ ~ 425 g/mol [56]. The polygycol (Sigma-Aldrich) was a random copolymer of 75% (by weight) ethylene- and 25% propylene-glycol (4:1 mole ratio), hydroxyl terminated with $M_w$ = 12.5 kg/mol. The perfluoropolyether (PFPE) was Fomblin Z25 from Solvay Solexis Inc. It is a linear copolymer of (-$CF_2$-$CF_2$-O-) and (-$CF_2$-O-) in the ratio of ~0.65, with terminal $CF_3$ groups. The molecular weight was 9.5 kg/mol (~ 95 repeat units per chain). The 1,1-(1,1,3-trimethyl-1,3-propanediyl)bis-cyclohexane (BCH) was obtained from Sigma Aldrich. The *p*-bis(phenylethyl) benzene (BPEB, $M_w$= 286 g/mol) was from Monsanto. The chemical structure of all materials analyzed herein are displayed in Table 1.

**RESULTS AND DISCUSSION**
**Non-associated liquids**

The interaction potential underlying the scaling is most obviously appropriate for non-associated glass-formers; accordingly, in Fig. 1 is shown the scaling of viscosity data [57] for octane, a prototypical van der Waals liquid. Isotherms for pressures over the range from 0.1 to ~370 MPa superpose when plotted vs. $TV^8$, demonstrating the extension of the thermodynamic scaling, previously applied to relaxation times, to $\eta$. The large value of $\gamma$ for octane reflects a strong volume contribution to the dynamics. We also note that the curvature of the octane data indicates an increase in the variation of $\eta$ with decreasing $TV^\gamma$; this differs from the usual behavior [17]. Since heretofore the scaling was investigated mainly for much larger $\tau$ (i.e., larger $\eta$), it is not clear if this is a true exception. Also in the figure are viscosity data for toluene [57], which are more nearly Arrhenius and characterized by a slightly smaller $\gamma = 7.5$. Interactions arising from the polarizable phenyl group are expected to emphasize the effect of temperature relative to that of volume. Thus, viscosities for BPEB, a model lubricant [58] having three benzene rings, superpose for an even smaller value of $\gamma = 4.8$ (Fig.2).

Viscosity master curves for two other non-polar liquids, BCH and PFPE, are also shown in Fig. 2. For the former $\gamma = 7.5$, which is quite similar to the behavior of the simpler organic liquids in Fig. 1. For the PFPE, however, superpositioning of $\eta$ requires a smaller scaling exponent = 6.0. Although not a large difference, this reduced volume-dependency is a consequence of the long chain character of PFPE; i.e., the greater concentration of intramolecular bonds. Eq.(1), which serves as the putative basis for the thermodynamic scaling, does not describe intramolecular bonding. A harmonic potential is more appropriate for changes in covalent bond lengths and angles arising from volume changes [59,60]. These intramolecular bonds thus serve to soften the potential.

This effect was seen in a recent mds of 1,4-polybutadiene by Tsolou et al. [30]. These authors employed an L-J 6-12 intermolecular potential in combination with harmonic backbone bonds to characterize the forces between chain segments [60]. The chain mobility data (both segmental and normal mode dynamics) for different $T$ and $V$ superposed according to eq.(2) with $\gamma = 2.8$ [30]. This is smaller than the value of 4 expected for the L-J potential ($r^{12} \sim V^4$), reflecting the influence of the intramolecular bonds. We illustrate this effect in Fig. 3, showing the intermolecular potential resulting from an L-J 6-12 interaction between non-bonded atoms in combination with harmonic chain backbone stretching

$$U(r) \sim \varepsilon\left[\left(\frac{\sigma}{r}\right)^{12} - 2\left(\frac{\sigma}{r}\right)^{6}\right] + k(r-l)^2 \quad (5)$$

In calculating the potential shown in the figure, we used average values for the three carbon-carbon bonds [60]: the L-J energy parameter $\varepsilon = 0.984$ kcal/mol, L-J length scale parameter $\sigma = 4.19$ Å, the backbone force constant $k = 821$ kcal/(mol Å$^2$), and backbone bond length $l = 1.46$ Å. While the asymptote at small $r$ indeed reflects the exponent 12 in eq.(5), in the vicinity of the local minimum the repulsive interaction is more gradual, describable by a power law $r^{-8.5}$. This same effect – a reduction in steepness of the repulsive potential due to a surfeit of intramolecular bonds – likely underlies the fact that superpositioning of dielectric relaxation times for polymers entails smaller $\gamma$ than the scaling exponent for molecular liquids [17,22].

Squalane is a saturated polybutene of modest molecular weight. Using the equation of state reported by Fandino et al. [61], a master curve for $\eta$ of squalane is obtained with $\gamma = 4.2$ (Figure 3). Mineral oil, an oligomeric hydrocarbon mixture of branched and linear alkanes and aliphatic and aromatic rings, exhibits behavior which is slightly less volume dependent, $\gamma = 4.0$. (Fig. 4). The exponents for both liquids are significantly smaller than the values for the van der Waals molecular liquids (Figs. 1 and 2) and the polymer, PFPE (Fig. 2). The latter demonstrates the effect of backbone structure. Although it is not possible to make a direct connection between chemical structure and $\gamma$, PFPE is non-polar and a very flexible polymer, having a facilely rotating ether linkage in every repeat unit. In the same manner that a flexible chain structure lowers the glass transition temperature (by mitigating constraints on the many-body dynamics [62]), the large value of $\gamma$ for PFPE (in comparison to the oligomers in Fig. 4 and general results for polymers [17,22] reveals a similar effect on the $V$-dependence of $\eta$.

To examine the connection between the interaction strength and the scaling exponent, we analyze data for a polar molecule (dipole moment = 2.4 D) [63]), dibutylphthalate (DBP). The measured temperature and pressure dependences of the liquid specific volume were fit to the Tait equation, yielding

$$V(T,P) = (1.0255 + 6.79\times10^{-4}T + 6.28\times10^{-7}T^2)\left[1 - 0.0894\ln\left(1 + \frac{P}{203.1\exp(-4.619\times10^{-3}T)}\right)\right] \quad (6)$$

with units of ml/g for $V$, and $P$ and $T$ in MPa and Celsius, respectively. As seen in Fig. 5, the viscosity data collapse for DBP onto a single master curve when plotted as a function of $TV^{3.2}$. This is a very small value of the exponent; in fact, DBP exhibits the lowest $\gamma$ found to date for any molecular glass-former lacking hydrogen bonds [17]. The inference is that the polarity and consequently enhanced interactions in DBP may soften the repulsive interactions, minifying the effect of volume on the dynamics relative to the influence of temperature. Also included in Fig. 5 are dielectric relaxation times for DBP measured at ambient and elevated pressures [64]. The relaxation times also scale for the same value of $\gamma = 3.2$; moreover, using eq.(4) with $G_\infty = 0.2$ GPa, the $\tau$ superpose onto the viscosity data. This value of $G_\infty$ is close to acoustic determinations of the high frequency shear modulus, = 0.33 GPa [65]. The combined superpositioning of $\eta$ and $\tau$ demonstrates that for a constant value of $\gamma$ the thermodynamic scaling applies over a broad dynamic range (almost 12 decades – encompassing the dynamic crossover [33]).

**H-bonded materials**

Hydrogen-bonded liquids are strongly associated, and thus volume exerts a weak effect on their dynamics [17,66,67]. Furthermore, if hydrogen bonds dissociation occurs with temperature (or in response to pressure [39,40,41]), the dynamics should depart from the thermodynamic scaling, since the chemical structure of the fluid is varying with $T$ and $P$. Nevertheless, relaxation times for at least some H-bonded liquids such as sorbitol ($\gamma = 0.13$ [21]) and glycerol ($\gamma = 1.6 \pm 0.2$ [26,27,28] do conform to eq.(2). Relaxation times for weakly H-bonded polymers such as polypropylene glycol with $M_w = 4$ kg/mol, in which only two of the 69 repeat units (the terminal end units) form hydrogen bonds, also superpose as a function of $TV^\gamma$ [68]. Generally, for polymers with terminal hydroxyl groups, the influence of volume (and thus $\gamma$) increases with molecular weight [69].

Fig. 4 shows the viscosities measured for a polyglycol with $M_w = 12.5$ kg/mol. The $\eta$ for three pressures collapse to a single curve for $\gamma = 2.3$. This value of the exponent is slightly less than for polypropylene glycol having lower $M_w$, for which $\gamma = 2.5$ [68]. When the molecular weight is sufficiently low, the increased H-bonding (more terminal hydroxyl groups) will cause a breakdown of the superpositioning. For polyglycol with $M_w = 0.4$ kg/mol, for example, we find (data not shown) that dielectric relaxation times do not superpose versus $TV^\gamma$.

The most common H-bonded liquid is water, which can form up to 4 hydrogen bonds per molecule. This number varies with temperature [70,71] and pressure [39], so that a breakdown of the thermodynamic scaling is expected. The inset to Fig. 5 shows the viscosity of water $T$ from 255K to 363K for pressures up to 900 MPa [72,73]. The data exhibit a minimum as a function of the specific volume, in accord with the temperature dependences of thermodynamic properties such as the density, heat capacity, and compressibility, which likewise go through a minimum in the vicinity of T ~ 300K [74]. This behavior of water is a consequence of the hydrogen bonds, specifically the geometrical constraints imposed by their formation. Directional interactions among water molecules give rise to an open H-bond structure, having a lower density and different properties than unassociated water molecules. The formation of this H-bond network is enhanced for a particular *T*-dependent volume [74] and this dependence of the chemical structure on thermodynamic conditions precludes a description of the viscosities as a unique function of $TV^\gamma$. The countervailing effects of pressure (volume) – greater congestion reducing mobility while disruption of H-bonding increases mobility – results in an extremely weak volume dependence of $\eta$. As seen in Fig. 5, the viscosity is almost constant as density is varied at constant temperature. The fact that temperature is the dominant control variable for the viscosity of strongly H-bonded liquids agrees with the general trend found for the relaxation time of various glass formers [17,66].

**Ionic liquids**

Ionic bonds are even stronger than H-bonds and hence tend to yield crystalline solids of high melting point. The molten liquid state of inorganic salts often requires temperatures above 500K, limiting experimental investigations. A feature inorganic salts have in common with water is a propensity to form network structures [75,76] and as discussed above, in general we cannot expect the thermodynamic scaling to apply to network-forming liquids. A recent mds of silica by De Michele et al. [77] demonstrates this; iso-diffusivities calculated for various thermodynamic conditions do not correspond to a constant value of $TV^\gamma$. Instead there are two regimes, due to the coupling between the density of silica and its ability to form bonds between Si and O atoms.

For ionic interactions weaker than in the inorganic salts, network formation is suppressed and conformance to the scaling may be observed. Salts in which the cation is bulky and asymmetric can dissociate at moderate temperatures to form a viscous liquid. Viscosity measurements have been carried out at elevated pressures on three ionic liquids for which PVT data are also available [44,45]. These materials have the typical structure of low temperature ionic liquids – a nitrogen-containing, organic cation and an inorganic anion, which poorly coordinate and thus melt at low temperature.

The viscosity data for each liquid superpose for low values of $\gamma < 3$ (Fig. 6). The dynamics primarily reflect the influence of temperature, consistent with strong intermolecular forces due to the electrostatic interactions. The smallest scaling exponent, $\gamma = 2.25$, is found for the [OMIM]BF$_4$. The tetrafluoroborate counter ion has a tetrahedral geometry and more compact size than the octahedral hexafluorophosphate: the F in BF$_4$ is 1.42Å from the central atom versus 1.65Å in PF$_6$ [78]. The result is a stronger effective negative charge (by about a factor of 2 [78]) and consequently stronger electrostatic interactions in [OMIM]BF$_4$. The exponent is larger for [OMIM]PF$_6$ in comparison to [BMIM]PF$_6$, which we can speculate is due to increased van der Waals interactions for the bulkier cation. This size effect is seen directly in the magnitude of the viscosity, which is about threefold larger for [OMIM]PF$_6$ than for [BMIM]PF$_6$. Although this first application of the scaling to ionic liquids is successful, certainly if would be of interest to determine if limits exist at extended densities or temperatures, as seen for H-bonded materials.

**CONCLUDING REMARKS**

Although the rationale for the thermodynamic scaling arose originally from a consideration of intermolecular forces and their relationship to local dynamics, obviously a spherical two-body repulsive potential (eq.(1)) can only crudely approximate the interactions between real molecules. Nevertheless, the superposition of relaxation times according to eq.(2) is an empirical fact, having been verified now for at least 50 molecular liquids and polymers [17,21,22,23,24,25,26,27,28,29,30,31,32]. The appeal of the analysis extends beyond the utility of the scaling to organize and extrapolate experimental data. The magnitude of the exponent $\gamma$ provides a direct link to the manner in which molecules negotiate a complex potential energy landscape, in which density-dependent local barriers exceed the available thermal energy. Molecular motion must proceed via many-body cooperative dynamics, which evolve with changing temperature and volume. A given value of $TV^\gamma$ corresponds to a constant average local arrangement (static structure factor) [18]. The scaling exponent quantifies the steepness of the intermolecular potential and thus the relative contribution of $T$ and $V$. It is for this reason that a connection exists between $\gamma$ and the Grüneisen parameter characterizing the anharmonicity of the potential [34,35].

Herein we extend the scaling to characterize the $T$- and $V$-dependences of the viscosity for 12 materials (Table 1). For the non-polar liquids (octane, toluene, BCH, and PFPE), $\gamma$ is large ($\geq 6$), consistent with earlier work showing the strong effect of volume on dielectric relaxation in supercooled van der Waals molecular liquids [79,80,81]. More polar interactions (as in DPEB and DBP herein) or more intramolecular bonding (squalane and mineral oil) reduces the influence of volume, thereby reducing $\gamma$. Polymers have a plethora of intramolecular bonds and thus generally weak volume effects [17,66]. For a polymer capable of H-bonding such as polyglycol, $\eta$ superpositioning requires an especially small value of $\gamma = 2.3$ due to the combined effects of strong interactions and many intramolecular bonds. More generally, the tendency of H-bonds to dissociate at higher $T$ or $P$ can cause deviations from eq.(2), since the structure of the material itself changes. This behavior is exemplified herein by water. Finally, we find that for ionic liquids, in which strong electrostatic forces are operative, the scaling exponent is invariably small (< 3). The differences in $\gamma$ among the three ionic liquids analyzed herein are not large, but they are consistent with the trend of stronger attractive interactions softening the hardcore potential, giving rise to larger $\gamma$.

**ACKNOWLEDGEMENTS**

This work was supported by the Office of Naval Research. Insightful comments by J. Budzien of Sandia National Laboratories are greatly appreciated.

Table 1. Materials and scaling exponents.

| | chemical name | structure | γ |
|---|---|---|---|
| octane | n-octane | 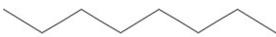 | 8.0 |
| toluene | toluene | 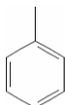 | 7.5 |
| BCH | 1,1-(1,1,3-trimethyl-1,3-propanediyl)bis-cyclohexane | 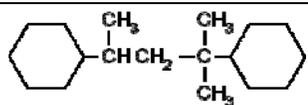 | 7.5 |
| PFPE | perfluoropoly(ethyleneoxide-ran-methyleneoxide) | $CF_3(-CF_2-CF_2-O-)_x(-CF_2-O-)_y-CF_3$ | 6.0 |
| BPEB | p-bis(phenylethyl)benzene | 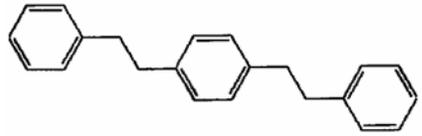 | 4.8 |
| squalane | 2,6,10,15,19,23-hexamethyltetracosane | 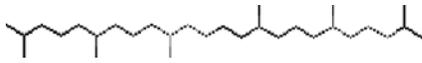 | 4.2 |
| mineral oil | linear and cyclic alkanes | NA | 4 |
| DBP | di-n-butylphthalate | 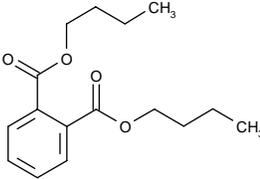 | 3.2 |
| polyglycol | poly(ethylene glycol–ran–propylene glycol) | $HO(-CH_2-CH_2-O-)_x(-CH_2-CH_2-CH_2-O-)_y-H$ | 2.3 |
| [OMIM]BF$_4$ | 1-methyl-3-octylimidazolium tetrafluoroborate | 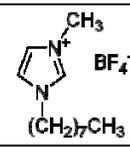 | 2.25 |
| [OMIM]PF$_6$ | 1-methyl-3-octylimidazolium hexafluorophosphate | 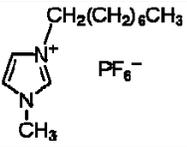 | 2.4 |
| [BMIM]PF$_6$ | 1-butyl-3-methylimidazolium hexafluorophosphate | 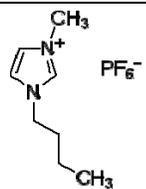 | 2.9 |

**Figure Captions**

Figure 1. Superpositioned viscosities of octane and toluene, with $\gamma = 8.0$ and 7.5 respectively. The $\eta$ were taken from ref. 57, with isotherms reported over the indicated pressure range and each symbol type representing a different measurement temperature.

Figure 2. Superpositioned viscosities of *p*-bis(phenylethyl)benzene ($\gamma = 4.8$), perfluoropolyether ($\gamma = 6.0$), and 1,1-(1,1,3-trimethyl-1,3-propanediyl)bis-cyclohexane ($\gamma = 7.5$). The range of pressures was from ambient to 610 MPa (BPEB), 1180 MPa (PFPE), and 300 MPa (BCH). Different symbols represent the different measurement temperatures of each isotherm (note that the viscosities of BPEB were shifted upward by one decade for clarity). Volumes, $V_R$, are relative values.

Figure 3. Intermolecular potential for 1,4-polybutadiene calculated as the sum of an L-J 6-12 intermolecular potential and a harmonic intramolecular bond stretching potential (eq.(5)). The parameters were taken from ref. 60, with average values used for the interaction parameters. The solid line is the fit to the repulsive energy in the vicinity of the minimum, the dashed line represents the limiting value of the repulsive energy at small *r*, and the dotted line is the bond stretching interaction.

Figure 4. Superpositioned viscosities of polyglycol ($\gamma = 2.3$), mineral oil ($\gamma = 4.0$), and squalane ($\gamma = 4.2$). The range of pressure for the measurements was from ambient to 1000 MPa (polyglycol), 1200 MPa (squalane), and 633 MPa (mineral oil). Different symbols represent different measurement temperatures.

Figure 5. Superpositioned viscosities of dibutylphthalate with $\gamma = 3.2$. The measurement pressure was from 0.1 to 1250 MPa at various temperatures as denoted by the symbol type. Dielectric relaxation times [64] for P ≤ 1610 MPa are included in the figure after multiplication by a factor of $2 \times 10^8$ Pa.

Figure 6. Arrhenius plots of the viscosity of water, indicating that temperature is the dominant control variable. The inset shows the negligible *V*-dependence, with a weak minimum related to the H-bond network structure. Collapse of the data when plotted versus $TV^\eta$ was not observed for any value of the exponent. The $\eta$ were taken from ref. 72 and 73.

Figure 7. Superpositioned viscosities of the ionic liquids: 1-methyl-3-octylimidazolium tetrafluoroborate ($\gamma = 2.25$), 1-methyl-3-octylimidazolium hexafluorophosphate ($\gamma = 2.4$), and 1-butyl-3-methylimidazolium hexafluorophosphate ($\gamma = 2.9$). The data were taken from ref. 44 ([BMIM]PF$_6$) and 45 (OMIM compounds). The pressure ranges from ambient to 224 MPa.

**Figure 1.**

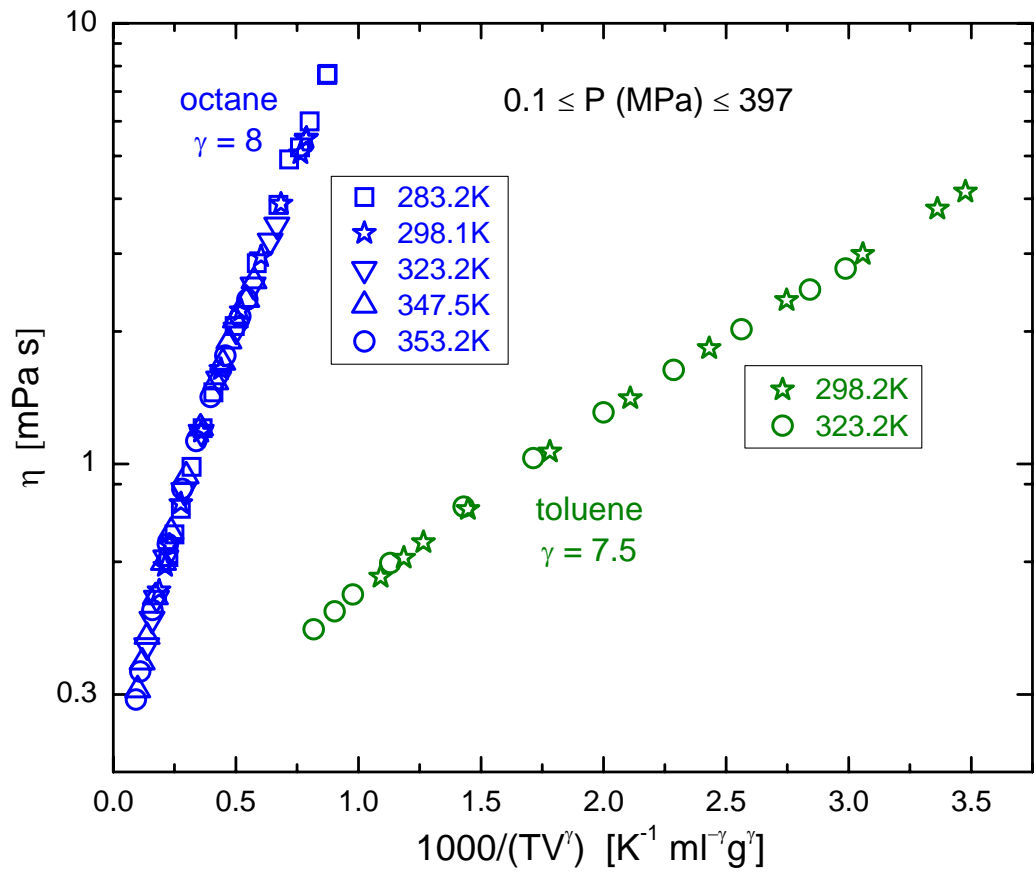

**Figure 2.**

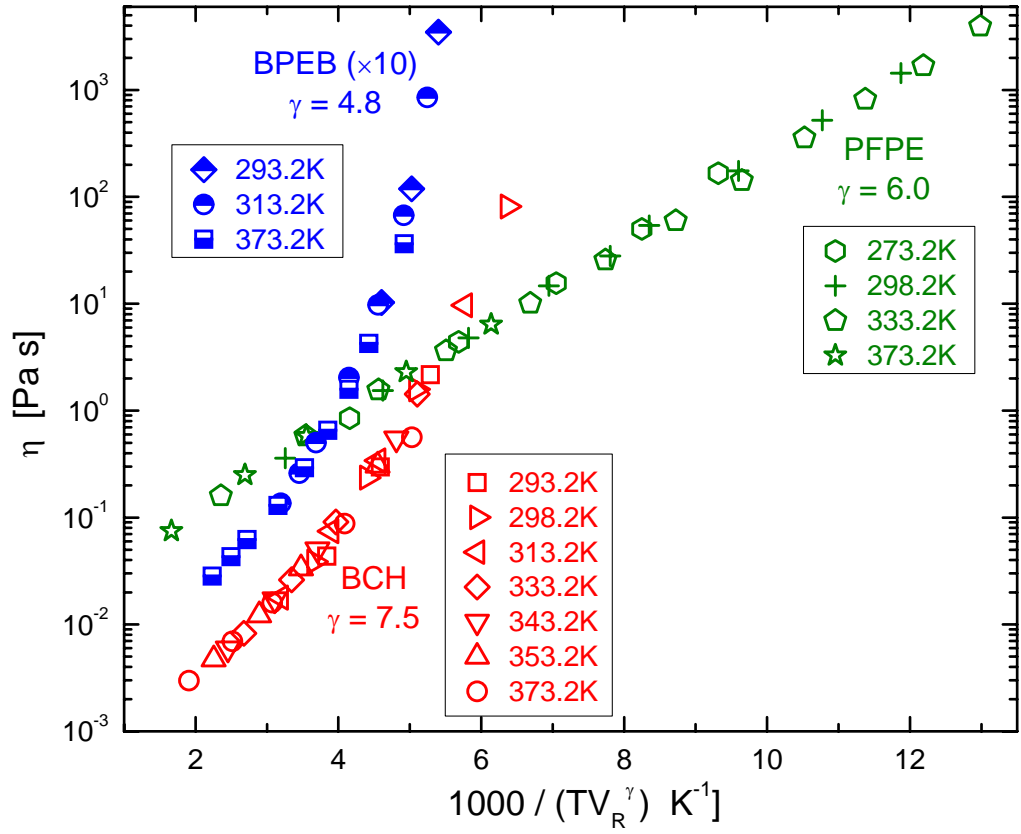

**Figure 3.**

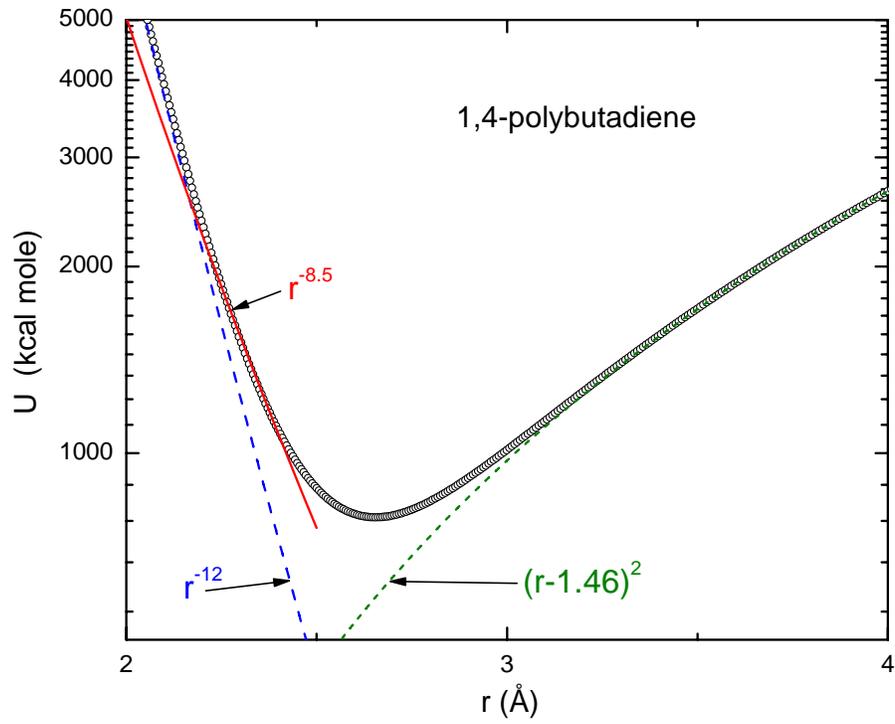

**Figure 4.**

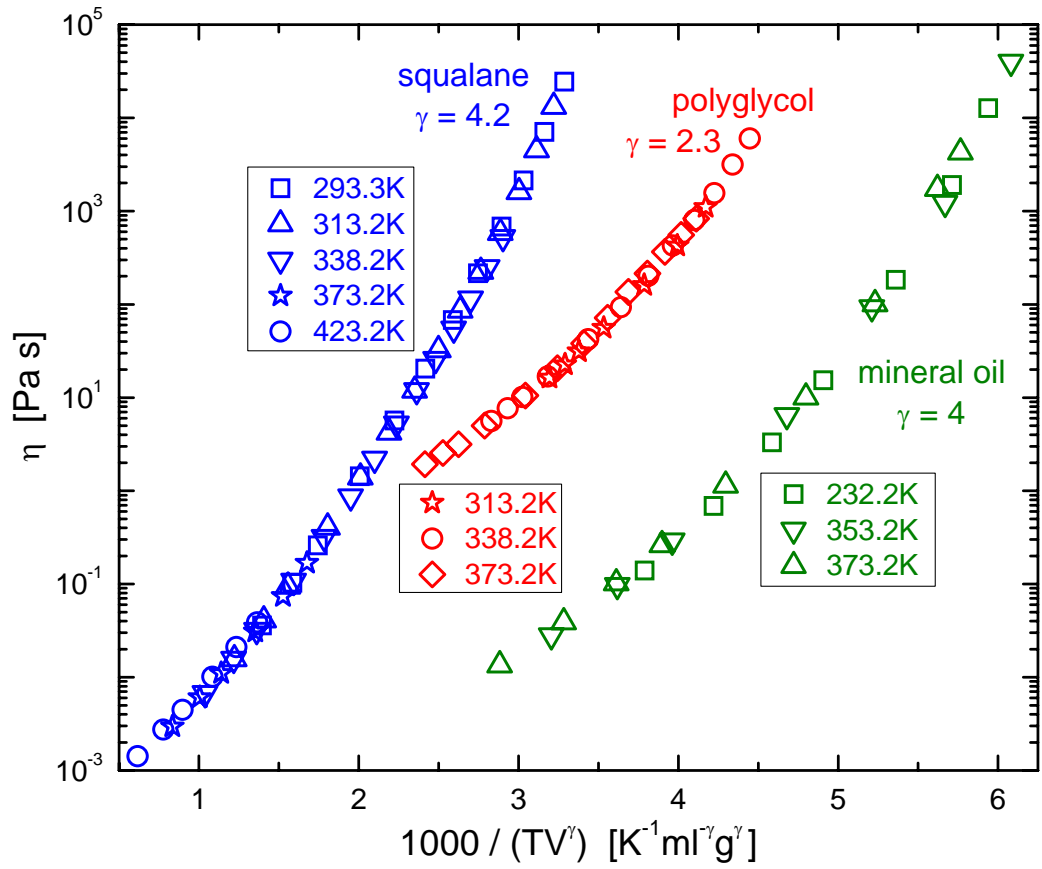

**Figure 5.**

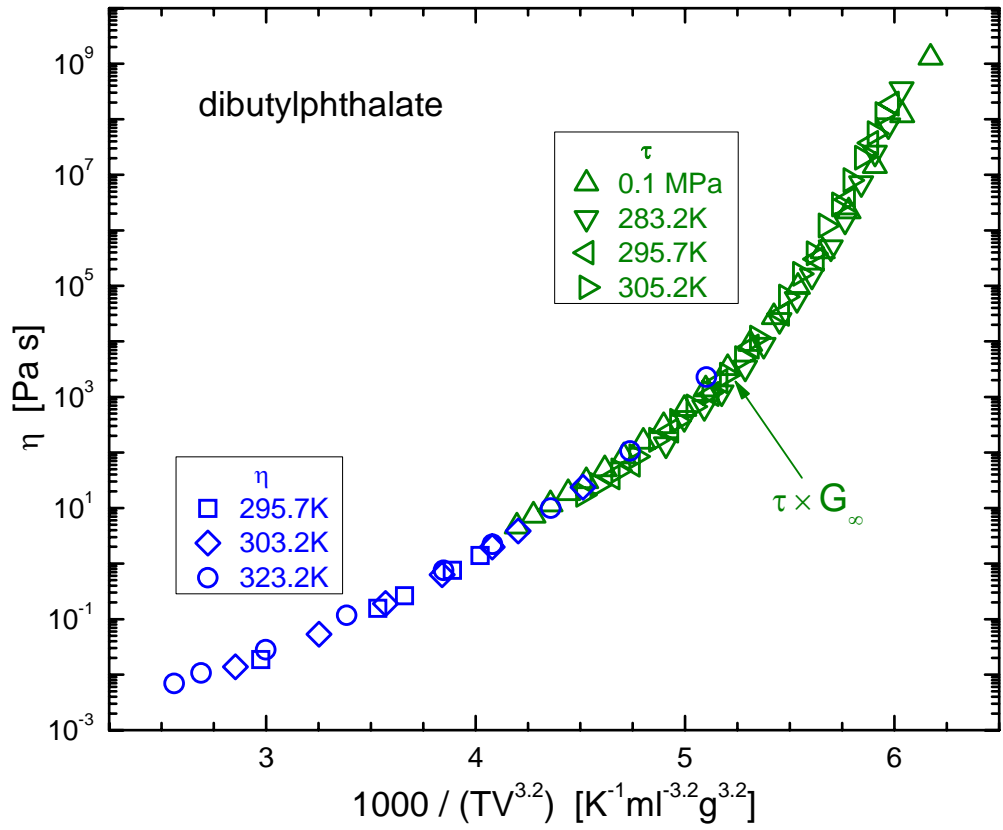

**Figure 6.**

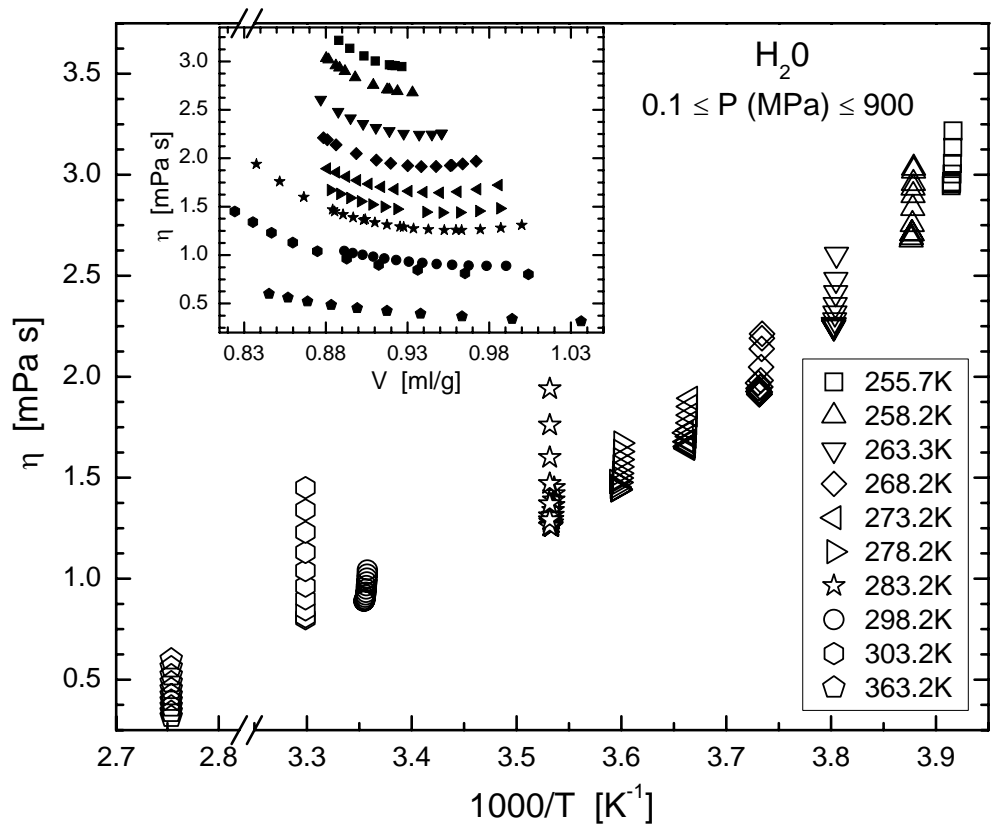

**Figure 7.**

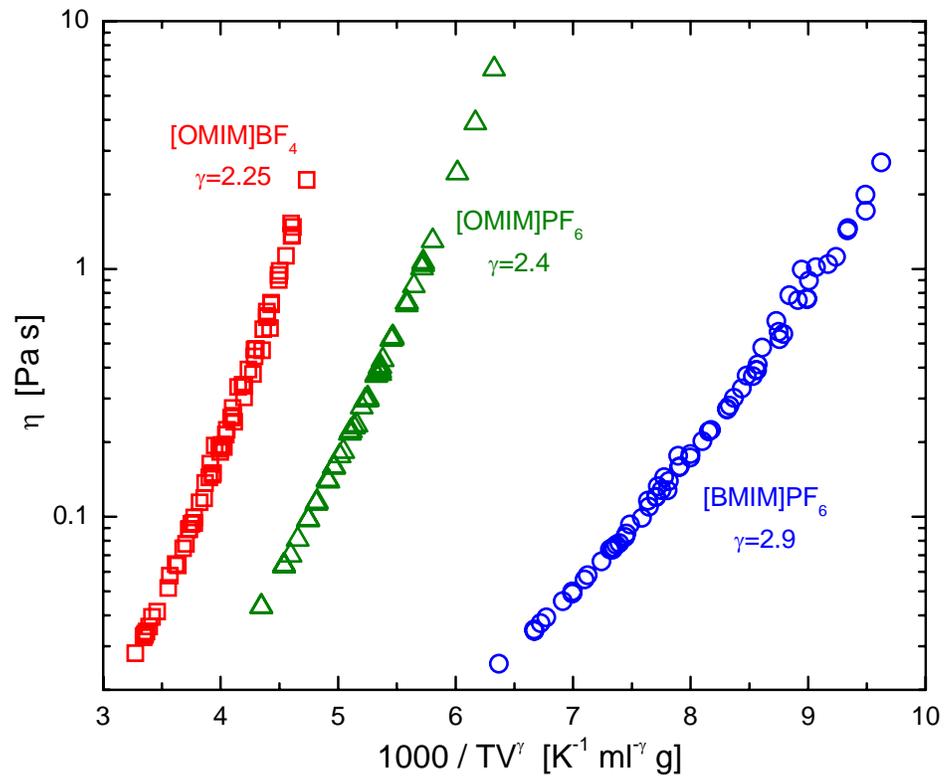